\documentclass[aps,floats,epsf,twocolumn]{revtex4}
 \usepackage{graphics}

\begin{document}

\title{Majorana mass, time reversal symmetry, and the dimension of space}

\author{Igor F. Herbut}
\affiliation{Department of Physics, Simon Fraser University, Burnaby, British Columbia, Canada V5A 1S6 \\
Max-Planck-Institut f\"ur Physik Komplexer Systeme, N\"othnitzer Strasse 38, 01187 Dresden, Germany}

\begin{abstract}
The Weyl fermions with a well defined chirality are known to demand that the dimension of space which they inhabit must be odd.
 It is shown here, however, that not all odd dimensional spaces are equally good hosts: in particular, an arbitrary number of chiral Weyl fermions can acquire a Majorana type of mass only in three (modulo eight) dimensions. The argument utilizes a) the precise analogy  between the Majorana mass term and the coupling of time-reversed Weyl fermions, and b) the conditions on the requisite time-reversal operator, which are implied by the real representations of Clifford algebras. In particular, it is shown that the latter allows only an even number of Majorana-massive Weyl fermions in seven (modulo eight) spatial dimensions. The theorem connects the observed odd number of neutrino flavors, the time reversal symmetry, and the dimension of our space, and strengthens the argument for the possible violation of the lepton number conservation law.

\end{abstract}
\maketitle

\vspace{10pt}

\section{Introduction}

Maybe the most obvious fact about our world is that it is three dimensional.\cite{comment0} Yet, neither classical nor quantum physics provide a very good reason why this needs to be so. On the contrary, modern approaches to the physics beyond the standard model based on the string theory, for example, typically require higher number of dimensions $d$ for internal consistency, and then their subsequent compactification down to $d=3$. The significance of three dimensions poses therefore a clear and important problem.\cite{karch}

 Here we point out a connection between the dimensionality of space and its ability to accommodate the Majorana mass for an arbitrary, and most importantly, an {\it odd} number of flavors of Weyl fermions. The connection is surprising in that it crucially involves the operation of time-reversal, which is a symmetry that {\it a priori} has little to do with any spatial characteristics, such as the dimension. We first show that the standard Majorana mass term in the Lagrangian for the Weyl fermions may be understood as a coupling  between the time-reversed states of the Weyl Hamiltonian.\cite{schrieffer} Focusing on odd spatial dimensions in which the notion of chirality is well-defined, we find that the time-reversal (TR) operator $T$, requisite for the construction of the Majorana mass, exists only in some of them: in $d=3+4 n$, $n=0,1,2...$. The reason for the absence of the TR operator for the Weyl Hamiltonian in every second odd dimension is related to its antilinearity.

 Furthermore, the square of the TR operator, when it exists, in dimensions $d=3$ and $d=7$ (modulo eight), depends on the dimension as $T^2= (-1)^{(d+1)/4}$. This has an important consequence for the mixing matrix between different flavors of Weyl fermions: the mixing matrix is {\it symmetric} only for $d=3$, whereas it is {\it antisymmetric} in $d=7$ (modulo eight). The latter condition in general implies double degeneracy of all finite eigenvalues, so that the number of massive Weyl fermions in this case, when $n$ is odd, can ultimately only be even. When the number of Weyl flavors is itself odd, this means that there is an odd number of Weyl fermions still being left massless. For three flavors in seven dimensions, for instance, this is illustrated further by a simple and direct calculation.

 The bottom line is that in order to have an arbitrary, and in particular, odd, number of Weyl fermions with {\it unconstrained} values of Majorana masses, the dimension of space has to be three, modulo eight. The ambiguity of eight in the result is a manifestation of the Bott periodicity in the theory of Clifford algebras,\cite{bott} upon which our proof ultimately relies. Some speculations about how to remove it will be offered.

The standard model of elementary particles famously contains three families of leptons, including the three neutrinos, which may be assumed to be chiral Weyl fermions. \cite{giunti} They are now believed to be endowed with small but finite masses. Whether the masses are of the usual Dirac type, or with the Majorana component is at present unknown, but the experiments sensitive to the character of the neutrino mass are on their way.\cite{avignon} The observation of the neutrinoless double beta decay, for example, which would imply a violation of the lepton number conservation which follows from the Majorana nature of the neutrino mass, would according to the theorem proved in this paper, if not quite explain, then to a certain degree rationalize the observed dimensionality of our space. The full explanation of the dimensionality along the direction taken here would demand an understanding of why the Majorana mass of the three neutrinos could be required for, as opposed to only being allowed in, our Universe.\cite{leptogenesis}

  The rest of the paper is organized as follows. In the next section we deconstruct the Weyl Hamiltonian, with the particular emphasis put on the basic assumptions behind it. In sec. III we lay out our main thesis: that implicit in the construction of the standard Majorana mass is the existence of the time-reversal operator that squares to minus one. In sec. IV it is shown that an operator with the requisite properties for the time-reversal exist only in the dimensions $8n+3$. The alternative point of view, which leads to the same conclusion but which is more in line with the standard group theory,\cite{georgi} is provided in sec. V. The important generalization to more than one flavor, with the properties of the mixing matrix in different dimensions, is given in sec. VI. The last section gives the conclusion. In the Appendix we discuss the special case of spatial dimensions $d=1+8n$, which allow a related, but physically different mass term that couples independent real Majorana-Weyl fermions, which can exist in these dimensions.

 \section{Weyl  Hamiltonian}

  We begin by establishing the basic terminology. Define the Weyl Hamiltonian as an irreducible, translational, and rotational invariant Hermitian operator, linear in momentum. The last requirement and the translational invariance together imply that
\begin{equation}
    H_W = \sum_{i=1} ^d \alpha_i p_i,
\end{equation}
where $p_i$ are the components of the momentum operator in $d$-dimensional space, $d\geq 2$, \cite{remark3}  and $\alpha_i$ are ``coefficients" that commute with the momenta $p_i$ and the coordinates $x_i$.  The rotational symmetry requires that both $p_i$ and $\alpha_i$ transform as components of vectors. It thus suffices\cite{footnote} that $\alpha_i$ obey the Clifford algebra $C(d,0)$: \cite{clifford}
\begin{equation}
[\alpha_i, \alpha_j ]_+ = 2\delta_{ij},
\end{equation}
where $[,]_s$ is the anticommutator for $s=+$, and commutator for $s=-$. The generators of rotations can be now constructed as the sum of the orbital angular momentum operator and the generators of spinor representation of the rotational group ($Spin(d)$) \cite{georgi}
\begin{equation}
    L_{ij} = p_i x_j - x_i p_j + \frac{i}{4} [ \alpha_i, \alpha_j]_-.
\end{equation}
 The coefficients $\alpha_i$  are therefore promoted by the rotational invariance into operators, and the Weyl Hamiltonian $H_W$ in fact acts in the Hilbert space
 \begin{equation}
    { \cal H} = {\cal H}_{orb} \otimes {\cal H}_{sp},
\end{equation}
where the coordinate and the momentum act in the first, orbital factor, and the operators $\alpha_i$ in the  second, spin space. The irreducibility of $H_W$ immediately implies that the representation of the Clifford algebra, and consequently of the $Spin(d)$, is $2^{(d-1)/2}$ dimensional, and that $d$ is {\it odd}. Instead of the irreducibility, one could have demanded that an operator $\beta$ which would anticommute with all $\alpha_i$ does not exist. This would guarantee that an addition of the Dirac mass term $\sim \beta$ to $H_W$ is impossible, or, in other words, that the masslessness of the Weyl particle is not accidental. Viewed either way, there are two inequivalent {\it irreducible} complex (and Hermitian) representations of $C(d,0)$ for odd $d$, which correspond to two possible chiralities of the Weyl Hamiltonian. Choosing the chirality causes then the Weyl Hamiltonian to break the symmetry of space inversion, as well known.

 \section{Majorana mass}

  Having defined the fundamental Weyl Hamiltonian, we next deconstruct the usual Majorana mass as the ``off-diagonal pairing" term, in terminology of condensed matter physics.\cite{schrieffer}
  It will be crucial that the Weyl particles are {\it fermions}, so consider the Lagrangian density for a {\it single} massless Weyl fermion:
  \begin{equation}
L_0 = \Psi^\dagger (i \partial _t + H_W)\Psi,
\end{equation}
where $\Psi= \Psi(\vec{x},t)$ and $\Psi^\dagger = \Psi^\dagger (\vec{x},t)= (\Psi^*)^T $ are independent $2^{(d-1)/2}$-component Grassmann (anticommuting) fields. To cast the Lagrangian into the form that would facilitate a natural addition of the mass term, we rewrite it using Nambu's particle-hole doubling\cite{schrieffer} as
  \begin{equation}
L_0 = \frac{1}{2} ( \Psi^\dagger , \tilde{\Psi}^\dagger ) (i \partial _t + \sigma_3 \otimes H_W) (\Psi^\dagger  , \tilde{\Psi}^\dagger )^\dagger,
\end{equation}
with $\tilde{\Psi} = U \Psi^*$, with the unitary matrix $U$ satisfying $U H_W ^* U^{-1} = H_W$. Summing the two decoupled blocks in the last equation recovers precisely Eq. (5). But, this construction provides the {\it unique} way to invert the sign of the Weyl Hamiltonian for the lower  (``hole") block in $L_0$, which, as will be seen, is necessary for the construction of the mass term for a single Weyl fermion. We now can recognize the combination $T_{sp}= U K$, with $K$ as the complex conjugation, as the spin part in the TR operator $T$ for the Weyl particle:
\begin{equation}
T = T_{orb} \otimes T_{sp},
\end{equation}
where $T_{orb}$ is the usual TR operator in the orbital space.\cite{gottfried} The Majorana mass is then simply
\begin{equation}
L_M = \frac{1}{2} ( \Psi^\dagger , \tilde{\Psi}^\dagger ) ((m_1 \sigma_1 + m_2 \sigma_2)\otimes 1 ) (\Psi^\dagger , \tilde{\Psi}^\dagger )^\dagger,
\end{equation}
and the massive Weyl particle is described by the Lagrangian $L= L_0 + L_M$.
The mass term breaks the global $U(1)$ particle number symmetry generated by $\sigma_3 \otimes 1$, and implies the relativistic energy spectrum
$\pm \sqrt{ k^2 + |m|^2}$, where $m=m_1 + i m_2$.

We will not be concerned here with the dynamical mechanism of the generation of the Majorana mass, which inevitably will be model dependent. Rather, we will focus on the structure of the Majorana mass itself. Evidently, the Majorana mass term is proportional to
\begin{equation}
 \Psi^\dagger U \Psi^* + c.c. = -\Psi^\dagger U^T \Psi^* + c.c.,
 \end{equation}
 with the minus sign on the right hand side reflecting the  Grassmann nature of the fields. The mass term thus either vanishes, or the matrix $U$ is antisymmetric, $U=-U^T$. Since $U$ also must be unitary, it follows that
 \begin{equation}
 -1 = U U^* = T_{sp}^2 = T^2,
 \end{equation}
 where in the last equation we used the fact that $T_{orb}^2 = 1$, always.\cite{gottfried}

 Two conditions are thus implicit in the construction of the  Majorana mass for a single Weyl fermion: a) the time-reversal operator for the Weyl Hamiltonian exists, and b) that the same time-reversal operator also has the usual negative sign when squared.

None of the above, of course, is new when $d=3$. The Clifford algebra elements then are the Pauli matrices, $\alpha_i = \sigma_i$, $i=1,2,3$, and the TR operator in the spin space is the familiar (and, up to a phase, unique) $T_{sp} = \sigma_2 K$, with the requisite value of the square: $T_{sp} ^2 = \sigma_2 \sigma_2 ^* =-1$. Writing the Lagrangian $L_M$ in terms of the Nambu components recovers precisely the textbook form of the Majorana mass.\cite{giunti} We demonstrate next that this construction is, surprisingly, possible only in every {\it fourth} odd dimension.\cite{comment5}

 \section{Time-reversal in different dimensions}

 \subsection{ Nonexistence in d=5}

 Before delving into the general proof, let us show that already in the next odd dimension of $d=5$ the operator $T_{sp}$ for the Weyl Hamiltonian simply {\it does not exist}. The irreducible representation of the Clifford algebra $C(5,0)$, modulo an overall sign and an  unitary transformation, is unique, and may be chosen to be, for example,
$\alpha_i = 1\otimes\sigma_i$, $i=1,3$,  $\alpha_2 = \sigma_2 \otimes\sigma_2$, $\alpha_4 = \sigma_1\otimes\sigma_2$, $ \alpha_5 = \sigma_3\otimes\sigma_2$, with $\alpha_i$ as real for $i=1,2,3$ and as imaginary for $i=4,5$. If $T_{sp} = U K$ and the Weyl Hamiltonian is to be even under time-reversal, all $\alpha_i$ must be odd. The matrix $U$ then needs to satisfy the conditions
\begin{equation}
[ U, \alpha_i ]_+ =  [ U,  \alpha_j]_- =0,
\end{equation}
for $i=1,2,3$, and $j=4,5$. But {\it the only} two linearly independent matrices that anticommute with the $\alpha_i$, $i=1,2,3$ are $\alpha_4$ and $\alpha_5$ themselves! Since these two  mutually anticommute, obviously there is no linear combination of them which would commute with both. In stark contrast to $d=3$, in $d=5$ the single-flavor Weyl Hamiltonian breaks {\it both} the symmetries of space-inversion and of time-inversion.

\subsection{Nonexistence in d=4n+1}

It is not too difficult to prove further that $T_{sp}$ cannot be found whenever the dimension of space is $d=4n+1$. The dimension of the irreducible representation of the relevant Clifford algebra $C(4n+1, 0)$ is $2^{2n}$, and it can be chosen so that $2n+1$ matrices are real, and the remaining $2n$ matrices are imaginary.\cite{comment1}  Let us choose $\alpha_i$, $i=1,... 2n+1$ as real, and with $i=2n+2,... 4n+1$ imaginary. The set of all different products of the first $4n$ matrices is easily shown to be linearly independent, and together with the unit matrix to provide a basis in the space of $2^{2n}$-dimensional matrices, which is $2^{4n}$-dimensional. The unitary part $U$ of the operator $T_{sp}$, would then have to commute with all imaginary $\alpha$-matrices, and anticommute with all real $\alpha$-matrices, in generalization of Eqs. (11). Let us then consider  the real $\alpha_1$ first. Half of the matrices in the above basis anticommute with it, whereas the other half commute. The operator $U$, if it exists, would then be in the first set of $2^{4n-1}$ linearly independent matrices. Let us then consider all the matrices {\it in that first set} that also anticommute with the second real matrix $\alpha_2$. Again, their number is a half of the original number, so we get a set of linearly independent candidates for $U$ of the size of $2^{4n-2}$. Each additional condition similarly halves the number of candidates, until we exhaust all but the very last imaginary matrix, $\alpha_{4n+1}$. Since the number of satisfied conditions is at this stage $4n$, there is  a {\it unique} matrix left by this construction which anticommutes with all the real $\alpha_i$ ($i=1,... 2n+1$) and commutes with all but the last one of the imaginary $\alpha_i$ ($i=2n+2, ... 4n$). That matrix is also easy to discern:
\begin{equation}
X= \prod_{i=2n+2}^{4n} \alpha_i.
\end{equation}
The same matrix X that satisfies the first $4n$ conditions cannot satisfy the last condition, however, since being a product of an odd number of $\alpha_i$ with $i\neq 4n+1$, instead of commuting, it {\it anticommutes}  with the last matrix:
\begin{equation}
[X, \alpha_{4n+1} ]_+ =0.
\end{equation}
Being unique in satisfying the first $4n$ conditions, we see that the sought operator $U$ does not exist in $d=4n+1$.

\subsection{Nonexistence in d=9: An alternative proof}

Since the case of nine dimensions provides the first non-trivial example of the claimed nonexistence of time-reversal operator in dimensions $d=4n+1$, here we provide an alternative proof, which relies on some of the properties of real representations of Clifford algebras.\cite{okubo, herbut1} Besides the standard representation consisting of five real and four imaginary matrices discussed earlier, the Clifford algebra $C(9,0)$ also allows an equivalent  sixteen-dimensional representation consisting of one real and eight imaginary matrices.  This follows from the fact that the real representation of $C(1,8)$ is sixteen dimensional.\cite{okubo} (See the Table 1. in \cite{herbut1}.) The unitary operator $U$ in the $I_t ^{spin}$ would therefore need to commute with all of the imaginary matrices in that representation, and anticommute with the single real matrix. But, since the real sixteen-dimensional representation of $C(0,8)$  is a ``normal" representation with unity as its sole Casimir operator,\cite{okubo} the only matrix that satisfies the former condition on $U$ is the unit matrix, which then obviously does not satisfy the latter condition.

In $d=9$, however, there is another and distinct possibility for the construction of the mass term, due to the existence of a purely real sixteen-dimensional representation of $C(9,0)$, which would represent a {\it real} Weyl, i. e.  ``Majorana-Weyl" fermion. We discuss this possibility in an equivalent, but somewhat simpler case of $d=1$ in the Appendix.

\subsection{ Existence in d=4n+3}

In odd dimensions $d=4n+3$, on the other hand, there is no difficulty in finding the unitary part of $T_{sp}$. Choosing $\alpha_i$ real for $i=1,...2n+2$ and imaginary for $i=2n+3, ... 4n+3$, the operator $U$ is unique and it equals
  \begin{equation}
  U= \prod_{i=2n+3}^{4n+3}  \alpha_i.
  \end{equation}
It is easy to confirm that $U$ now satisfies all of the $4n+3$ desired conditions, and that it anticommutes (commutes) with all real (imaginary) $\alpha$-matrices.

Finally, the last equation implies that, when it exists,  $T_{sp} ^2 = (-1) ^{n+1}$. Comparing with Eq. (10) we see that
the Majorana mass term for a single Weyl flavor survives only in dimensions $d=3 + 8 n$, with $n=0,1,2,..$.

\section{ Relation to reality of spinor representations}

The conclusion that there is no time-reversal operator in spatial dimensions $d=4n+1$ may appear particularly surprising to a reader knowledgeable about the reality properties of the spinor representations of the rotational group.\cite{georgi} Namely, the generators of the spinor representation of $SO(d)$,
  \begin{equation}
  G_{ij} = \frac{i}{4} [ \alpha_i, \alpha_j]_-,
  \end{equation}
  with $\alpha_i$, $i=1,...d$, satisfying Clifford algebra $C(d,0)$, allow one to find a {\it unique} matrix $R$ so that
  \begin{equation}
  G_{ij} = - R G_{ij} ^* R^{-1},
  \end{equation}
  with the matrix $R$ being either symmetric (``real" representations) or antisymmetric (``pseudo-real" representations). It follows that the spinor representations of $SO(p + 8k)$ are real when $p=1,7$ and pseudoreal when $p=3,5$.\cite{georgi}

The above condition can be understood precisely as the oddness of the generators $G_{ij}$ under the antilinear operator $A= R K$. If the matrix $R$ is
unitary, one further finds that the antilinear operator is such that either $A^2=1$ for real, or $A^2 =-1$ for pseudoreal representations. If one would identify $A$ as the operator of time-reversal, the statement would be that the generators of the spinor representation are odd under the time-reversal operator, which is unique, and with the square that depends on the dimension of space. The oddness of the rotational generators then implies that the rotations themselves are even under time-reversal, just as they are in the orbital space.

Our identification of the time-reversal operator in dimensions $d=3$ and $d=7$ (modulo eight), is evidently in perfect agreement with the above reality properties of $SO(3)$ and $SO(7)$, where, for each of these two dimensions, we identified a unique time-reversal operator under which $\alpha_i$ and therefore $G_{ij}$ are odd, and which has a square with the required sign. In $d=5$ and $d=9$ (modulo eight), on the other hand, we could not find any antilinear operator which would anticommute with $\alpha$-matrices and thus represent the operation of time-reversal. This conclusion, although at first it may appear otherwise, is in fact not in collision with the established facts about the reality of spinor representations in the dimensions in question. This is because the oddness of all $\alpha$-matrices is certainly a sufficient, but {\it not also a necessary requirement} for the desired oddness of the rotational generators. Since the generators are products of two $\alpha$-matrices and the imaginary unit, obviously even if all the $\alpha$-matrices were even under some antilinear operator, the rotational generators would still be odd. This is exactly the case in $d=5$ and $d=9$, as we now show. In the representation of $C(5,0)$ right above Eq. (11), the requisite antilinear operator is unique, and
\begin{equation}
A=i\alpha_4 \alpha_5 K.
\end{equation}
In $d=5$ therefore, $A^2 = -1$, and the spinor representation of $SO(5)$ is indeed pseudoreal. In the sixteen-dimensional representation of $C(9,0)$
with only one matrix ($\alpha_1$) real, and the remaining eight imaginary, mentioned earlier, the desired antilinear operator is the simplest to write:
\begin{equation}
A= \alpha_1 K,
\end{equation}
so $A^2 = 1$, and $SO(9)$ is real. In either case, and in contrast to $d=3$ and $d=7$, while in dimensions $d=5$ and $d=9$ the generators of rotations in spinor  representation are odd under a unique antilinear operator, their building blocks, the $\alpha$-matrices, are not odd but {\it even} under the same operation. Since the Weyl Hamiltonian is linear in both the momentum and in the $\alpha$-matrices, its time-reversal invariance demands a {\it stronger} condition than just oddness of the rotational generators: the oddness of the underlying Clifford algebra itself. It is this stronger condition that cannot be met in dimensions $d=5$ and $d=9$ (modulo eight).

Let us finally connect the antilinear operator $A$ to the the symmetry of the Weyl Hamiltonian, and eventually to the existence of the Majorana mass. We have concluded above that there are two options:
  \begin{equation}
  [A,\alpha_i ]_s =0,
  \end{equation}
  with the sign $s=\pm$, which then imply
  \begin{equation}
  [A,H_W]_{-s} =0.
  \end{equation}
  Here, $s=+$ corresponds to dimensions $d= 3,7$, and $s=-$ to dimensions $d=1,5$. If the Weyl equation is
  \begin{equation}
  i\partial_t \Psi = H_W \Psi
  \end{equation}
  then,
  \begin{equation}
  - i\partial_t (A \Psi) = s H_W (A \Psi).
  \end{equation}

  If $s=-$, the spinor $A\Psi$ satisfies exactly the same equation as the original spinor. We may interpret $A$ as the operator of the particle-hole symmetry. Since the operator $A$ is the unique antilinear operator under which the spinor group generators are odd, there is no operator that would correspond to the time-reversal symmetry in these dimensions. We may write the equation for the combined spinor $\Phi = (\Psi, A \Psi)^T$ as
  \begin{equation}
  i\partial_t \Phi = ( \sigma_0 \otimes H_W)   \Phi.
  \end{equation}
  Since any matrix commutes with the two-dimensional unit matrix $\sigma_0$, and by construction there is no matrix that anticommutes with $H_W$,  there is evidently no
  matrix that anticommutes with $\sigma_0 \otimes H_W$, i. e. no mass term is possible.

  If $s=+$, on the other hand, the situation is rather different, since the spinor $A\Psi$ then satisfies the same equation as $\Psi$, but with the {\it time axis reversed}. $A$ is therefore nothing but the time-reversal operator, whereas the Weyl equation now does not possess the particle-hole symmetry. The combined spinor $\Phi$ now satisfies the equation
 \begin{equation}
  i\partial_t \Phi = ( \sigma_3 \otimes H_W )  \Phi,
  \end{equation}
in which we traded the reversal of the time-axis for the lower component for the sign-change on the right-hand side. Because of the appearance of the Pauli matrix $\sigma_3$ in the last equation, there are now two matrices that anticommute with the $\sigma_3 \otimes H_W$: $\sigma_1 \otimes 1$ and $\sigma_2 \otimes 1$. These are the possible (Majorana) mass terms.

  We conclude that the irreducible Weyl equation in different dimensions
  is either odd under time-reversal, or odd under particle-hole conjugation. Since it is by its very nature always odd under parity,
it is invariant under the combined operation of particle-hole conjugation, parity, and time-reversal, in any dimension, just as one would expect from its Lorentz invariance. \cite{greenberg}.

\section{ Flavors and their mixing}

The mass term for $N>1$ flavors of Weyl fermions with equal chirality now generalizes into
\begin{widetext}
\begin{equation}
L_M = \frac{1}{2} ( \Psi^\dagger , \tilde{\Psi}^\dagger ) (\sigma_1 \otimes (m O + m^* O^\dagger) \otimes 1 ) +
i \sigma_2\otimes (m O - m^* O^\dagger) \otimes 1 )(\Psi^\dagger , \tilde{\Psi}^\dagger )^\dagger,
\end{equation}
\end{widetext}
where $O$ is the $N$-dimensional mixing matrix which acts in the flavor space, and $\Psi$ stands for $N$ different Weyl fields.
Finiteness of the Majorana mass term now implies that
$-O^T \otimes U^T = O \otimes U $, or equivalently
\begin{equation}
O^T = - (T^2) O.
\end{equation}
The mixing matrix $O$ is therefore symmetric in $d=3$ (modulo eight), as well known. However, it is antisymmetric in the second set of dimensions which allow the time-reversal operator $T$, namely $d=7$ (modulo eight). In the latter case the Majorana mass spectrum becomes severely restricted, as we now show.

Squaring the quadratic form in the Lagrangian now yields the spectrum,
 \begin{equation}
 \omega_i = \pm \sqrt{ k^2 + |m|^2 o_i },
 \end{equation}
where $o_i$ $i=1,2,...N$ are the eigenvalues of the positive matrix $O O^\dagger$. Obviously,
\begin{equation}
\prod_{i=1} ^N o_i = \det (O O^\dagger) = |\det O|^2,
\end{equation}
where we used the fact that $\det O = \det O^T$ in the last equality. On the other hand, Eq. (26) then implies that it is also true that
\begin{equation}
\prod_{i=1} ^N o_i =  [-(T^2)]^N  \prod_{i=1} ^N o_i,
\end{equation}
and at least one eigenvalue $o_i$ must vanish whenever $T^2 = 1$, as in $d=7$ (modulo eight),  and the number of Weyl flavors $N$ is odd.

The above conclusion is a consequence of the useful decomposition\cite{youla} of an antisymmetric matrix: there exists a transformation $O= W Q W^T $ with the matrix $W$ being unitary, so that the matrix $Q$ is block-diagonal, with each block as being either zero, or as the  two-dimensional matrix $q_i \sigma_2$, with complex $q_i$. The eigenvalues of the matrix $O O^\dagger$ are then $o_i = |q_i|^2$, each doubly degenerate, or $o_i = 0$. If the number of flavors $N$ is odd, an odd number of Weyl fermions remains massless, while the rest are pairwise degenerate.  If the mixing matrix $O$ is symmetric, on the other hand, the eigenvalues $o_i$ are unrestricted, and their degeneracies are only accidental.

\subsection{ Three flavors}

Since in nature there exist three types of neutrinos,
let us examine the case of $N=3$ more closely. A general three-dimensional mixing matrix can be written as
  \begin{equation}
  O= a + \sum_{i=1}^3 b_{i} J_i  + \sum_{i,j} c_{ij} T_{ij},
  \end{equation}
  where $J_i$ $i=1,2,3$ are spin-one angular momentum operators, and $T_{ij} = (1/2) [J_i, J_j ]_+ - (2/3) \delta_{ij}$  are the components of the
  antisymmetric tensor.\cite{gottfried} In the adjoint representation, $[J_i]_{jk} = -i \epsilon_{ijk}$, and all three angular momentum operators are antisymmetric matrices, whereas the components of the tensor operator are all symmetric matrices. In $d=7$ (modulo eight) then $a=c_{ij} \equiv 0$, to ensure the antisymmetry of the mixing matrix. In this case
  \begin{equation}
  O O^\dagger = \sum_{i,j=1}^3 b_i b_j ^* J_i J_j
  \end{equation}
  and the eigenvalues are found easily to be $\lambda_{1,2} = \sum_i |b_i | ^2 $, and $\lambda_3=0$. In contrast, in $d=3$ (modulo eight) the mixing matrix $O$ is symmetric, and it is then $b_i \equiv 0$ in Eq. (20). The remaining six linearly independent terms in Eq. (30) allow then an unconstrained mass spectrum.

\section{Conclusion}

 We have shown that an odd number of (Majorana) massive Weyl fermions can be accommodated only in three, modulo eight, dimensions.
All other dimensions are forbidden, but for different reasons: a) first, and as known already, in even dimensions the Weyl Hamiltonian is reducible, b) in five (modulo four) dimensions the Weyl Hamiltonian breaks the time reversal symmetry, so that the Majorana mass term is impossible, and finally c) in seven (modulo eight) dimensions the TR operator for the Weyl Hamiltonian exists, but it has a positive square, which implies an exact zero mode of the (antisymmetric) mixing matrix, and double degeneracy of the rest of the spectrum.

Assuming that nature avoids unnecessary masslessness,\cite{comment2}  but for some, at present, not well understood reason favors having Weyl fermions in odd number of copies, implies that the space must be three (modulo eight) dimensional. The ambiguity of eight is inherent to our Clifford-algebraic argument, and it is a mathematical consequence of the Bott periodicity.\cite{bott} It thus seems likely that some arguments beyond the mere consistency requirements would be required to remove it. For example, if one subscribes to the superstring or the M-theory, the number of spatial dimensions before compactification is nine and ten, respectively. Since the next allowed dimension of space in the present calculation is {\it eleven}, these theories would have no choice but to be compactified down to precisely three dimensions in order to allow three Majorana massive Weyl fermions. Taken more conservatively, our result provides a reason to hope that, given that we do live in three dimensions, nature would not miss the rare opportunity to provide the observed three neutrinos with the Majorana type of mass.

\section{Acknowledgements}

This work was supported by the NSERC of Canada. The author is grateful to D. Brody, J. Cayssol, K. Imura, R. Jackiw, C. Mudry, F. Noguira, M. Rios, G. Semenoff, and G. Volovik  for useful discussions and correspondences.

\section{Appendix: Majorana-Weyl fermions in d=1+8n}

  We have shown that the Majorana mass term that would be analogous to the one in $d=3$ cannot be written in other dimensions, and in particular not in
  dimensions $d=1+8n$. On the other hand, one can easily write down a real solution of the Dirac equation in these dimensions. Consider the simplest example in $d=1$:
  \begin{equation}
  i\partial_t \Psi = (\sigma_3 p + m\sigma_2 ) \Psi,
  \end{equation}
  with the momentum operator $p=-i \partial_x$. Since the imaginary unit can be canceled, the spinor $\Psi$ is real. Is that not in contradiction with our claim  \cite{jackiw}? 

    When $m=0$ the above equation decouples into two equations for the independent components of the spinor $\Psi = (\Psi_1, \Psi_2)^T$:
  \begin{equation}
  \partial_t \Psi_n = (-1)^n \partial_x \Psi_n,
  \end{equation}
  where $n=1,2$. Each equation can be understood as the Weyl equation for a one-component fermion, which, however, is {\it real}. In dimensions $d=1+8n$ the irreducible Weyl equation happens to have purely real representation, and therefore in these dimensions we have the unique possibility of a real ``Majorana-Weyl"  fermion.

  The mass term in this case is thus not the coupling of the (complex) Weyl fermion to its time-reversed copy, as it was in $d=3+8n$, but rather the coupling of one (real) Majorana-Weyl fermion to another, independent Majorana-Weyl fermion, with opposite chirality. Indeed, the two equations for the real $\Psi_n$, $n=1,2$ cannot be recombined into a single irreducible Weyl equation for the single-component complex Weyl fermion $\Phi = \Psi_1 + i \Psi_2$. Instead, put together they are equivalent to
  \begin{equation}
  i\partial_t  \Phi = p \Phi^*.
  \end{equation}

  The standard {\it linear}  irreducible (one-component) Weyl equation for the complex fermion in $d=1$ is thus not equivalent to the system of two Weyl equations for two real Majorana-Weyl fermions of opposite chiralities. This reconciles our demonstration that while the former does not allow the introduction of the mass term, the latter obviously does. The situation repeats itself, by Bott periodicity in higher dimensions, and in particular in $d=9$.\cite{gliozzi}


\begin{thebibliography}{99}
\bibitem{comment0} We will always refer here to the dimensions of space alone, and assume there is a single dimension of time.
\bibitem{remark3} Since we will view here the rotational invariance as the {\it raison d'etre} for the Weyl Hamiltonian, we focus on spatial
dimensions $d \geq 2$, in which the notion of rotation is sensible. Nevertheless, the case of $d=1$  will turn out to be equivalent to $d=9$, which will be explicitly studied in sec. IV c, by the Bott periodicity.
\bibitem{footnote} This is not a necessary condition, however, and in $d=3$ $\alpha_i$ can also close $SU(2)$ algebra for spin $s$. For any $s > 1/2$, however, the spectrum will then exhibit more than one characteristic velocity, and the Lorentz invariance will be absent, unlike when $\alpha_i$ close the Clifford algebra.
\bibitem{clifford} $C(p,q)$ is defined as a set of $p+q$ mutually anticommuting generators, $p$ ($q$) of which square to $+1$ ($-1$).
\bibitem{comment5} It may be more common, but equivalent to our point of view, to understand the operator $\sigma_2 K$ as charge conjugation, under which the chirality of the Weyl Hamiltonian becomes reversed. In $d=3$, $L_0$ then manifestly violates the charge conjugation while respecting the time reversal symmetry. (See also section V.)
\bibitem{comment1} This is because the real representation of the related Clifford algebra $C(2n+1,2n)$ is $2^{2n}$-dimensional. Upon multiplication of the latter $2n$ matrices by the imaginary unit we obtain the desired complex representation of $C(4n+1,0)$. See, Ref. \cite{okubo}, or the summarizing Table 1. in \cite{herbut1}.
\end{thebibliography}
\end{document}